\documentclass[twocolumn,showpacs,preprintnumbers,superscriptaddress,amsmath,amssymb,nofootinbib]{revtex4}

\usepackage{graphicx}
\usepackage{dcolumn}
\usepackage{bm}
\usepackage{amsmath, amssymb}

\usepackage[normalem]{ulem}  
\usepackage[dvipdfmx]{color} 

\renewcommand\sout{\bgroup \color{red} \ULdepth=-.5ex \ULset}

\newcommand{\Psfig}[2]{\includegraphics[width=#1]{#2}}
\newcommand{\PsfigII}[2]{\includegraphics[scale=#1]{#2}}

\def\Eglab{E_{\gamma}^{\rm lab} }
\def\deutwf{\tilde{\varphi} }
\def\eprime{\eta ^{\prime} }

\def\Kaellen{K\"{a}llen }

\def\mev{\text{ MeV}}
\def\gev{\text{ GeV}}

\def\mcb{~ \mu \text{b}}

\begin{document}

\preprint{}

\title{Theoretical study of photoproduction of an $\bm{\eta ^{\prime}
    N}$ bound state on a deuteron target \\ with forward proton emission}

\author{Takayasu Sekihara} 
\email{sekihara@post.j-parc.jp}

\altaffiliation[The present address: ]{Advanced Science Research
  Center, Japan Atomic Energy Agency, Shirakata, Tokai, Ibaraki,
  319-1195, Japan}

\affiliation{Research Center for Nuclear Physics (RCNP), Osaka
  University, Ibaraki, Osaka, 567-0047, Japan}

\author{Shuntaro Sakai} 
\altaffiliation[The present address: ]{Research Center for Nuclear
  Physics (RCNP), Osaka University, Ibaraki, Osaka, 567-0047, Japan}
\affiliation{Department of Physics, Kyoto University,
  Kitashirakawa-Oiwakecho, Kyoto 606-8502, Japan}

\author{Daisuke Jido}
\affiliation{Department of Physics, Tokyo Metropolitan University,
  Hachioji 192-0397, Japan}

\date{\today}

\begin{abstract}

  Possibilities of observing a signal of an $\eprime n$ bound state
  are investigated by considering photoproductions of the $\eta$ and
  $\eprime$ mesons on a deuteron target with forward proton emission.
  For this purpose, we take the $\eprime n$ interaction from the
  linear sigma model with a coupling to $\eta n$, in which an $s$-wave
  $\eprime n$ bound state can be dynamically generated, and we fix the
  $\gamma p \to \eta p$ and $\eprime p$ scattering amplitudes so as to
  reproduce the experimental cross sections with forward proton
  emission.  By using these $\gamma p \to \eta ^{( \prime )} p$ and
  $\eta ^{( \prime )} n \to \eta ^{( \prime )} n$ amplitudes, we
  calculate cross sections of the $\gamma d \to \eta n p$ and $\eprime
  n p$ reactions with forward proton emission in single and $\eta
  ^{(\prime )}$-exchange double scattering processes.  As a result, we
  find that the signal of the $\eprime n$ bound state can be seen
  below the $\eprime n$ threshold in the $\eta n$ invariant mass
  spectrum of the $\gamma d \to \eta n p$ reaction and is comparable
  with the contribution from the quasifree $\eprime$ production above
  the $\eprime n$ threshold.  We also discuss the behavior of the
  signal of the $\eprime n$ bound state in several experimental
  conditions and model parameters.
  
\end{abstract}


\maketitle

\section{Introduction}

The clarification of properties of the $\eprime$ meson is one of the
important topics in hadron physics.  Its anomalously heavy mass, known
as the $\text{U}_{\rm A}(1)$ problem~\cite{Weinberg:1975ui}, can be
explained by the fact that the $\text{U}_{\rm A}(1)$ symmetry is
explicitly broken by quantum anomaly in quantum chromodynamics
(QCD)~\cite{Adler:1969gk, Bell:1969ts, Bardeen:1969md} and the
$\eta^{\prime}$ meson is not a Nambu-Goldstone boson associated with
the chiral symmetry breaking~\cite{'tHooft:1976up, 'tHooft:1976fv,
  Witten:1979vv, Veneziano:1979ec}.  It is also important to emphasize
that the U$_{\rm A}$(1) anomaly is not the only source of the mass of
the $\eta^{\prime}$ meson, but the SU(3) chiral symmetry is necessarily
broken for the anomaly to affect the $\eta^{\prime}$ mass
spectrum~\cite{Lee:1996zy, Jido:2011pq}.

One of the recent interests in the $\eta^{\prime}$ meson is its
in-medium properties~\cite{Pisarski:1983ms, Bernard:1987sg,
  Kunihiro:1989my, Kapusta:1995ww, Tsushima:1998qp, Costa:2002gk,
  Nagahiro:2004qz, Bass:2005hn, Nagahiro:2006dr, Jido:2011pq,
  Nagahiro:2011fi, Nanova:2012vw, Nanova:2013fxl, Sakai:2013nba},
especially in the context of partial restoration of chiral symmetry in
nuclear matter~\cite{Jido:2011pq}.  As mentioned above, the
$\eta^{\prime}$ mass is closely related also to the chiral symmetry
breaking. In the nuclear medium, chiral symmetry is considered to be
partially restored with 30\% reduction of the magnitude of the quark
condensate at the saturation density~\cite{Suzuki:2002ae}.  Thus, the
$\eta^{\prime}$ mass is expected to be reduced in the nuclear
matter. A simple estimation based on partial restoration of chiral
symmetry has suggested about 100 MeV reduction of the $\eta^{\prime}$
mass at the saturation density~\cite{Jido:2011pq} as seen in the
chiral effective model calculations by the NJL
model~\cite{Costa:2002gk} and the linear sigma
model~\cite{Sakai:2013nba}.  The strong mass reduction in nuclear
matter provides a strong attractive scalar potential for the
$\eta^{\prime}$ meson in finite nuclei. This has stimulated
experimental and theoretical studies of search for $\eta^{\prime}$
bound states in nuclei~\cite{Itahashi:2012ut, Nagahiro:2012aq}.

According to the linear sigma model, if the dynamical chiral symmetry
breaking plays an important role for the mass generation of a hadron,
the hadron should have strong coupling to the $\sigma$ field.
Recalling that (a part of) the nucleon ($N$) mass is generated by the
chiral symmetry breaking and the $\sigma$ exchange provides a strong
attraction for the $N N$ interaction in the isoscalar-scalar channel,
one expects a similar attraction in the $\eta^{\prime}N$ interaction
and a possible two-body bound state of
$\eta^{\prime}N$~\cite{Sakai:2013nba}.  Thus, the interaction between
$\eprime$ and $N$ is a key to investigate properties of the $\eprime$
meson. The $\eta^{\prime} N$ interaction was investigated in, {\it
  e.g.}, the chiral effective models~\cite{Kawarabayashi:1980uh,
  Borasoy:1999nd, Oset:2010ub}.  A possibility to form an $\eprime N$
bound state was pointed out in the linear sigma model in
Ref.~\cite{Sakai:2013nba}.  An experimental signal of the $\eprime N$
bound state was implied in Ref.~\cite{Moyssides:1983}, where they
measured the $\pi ^{-} p \to \eprime n$ cross section just above the
$\eprime n$ threshold.  Production experiments of the $\eprime$ meson
in other reactions, such as $\gamma p \to \eprime
p$~\cite{Dugger:2005my, Williams:2009yj, Crede:2009zzb,
  Sumihama:2009gf} and $p p \to \eprime p p$~\cite{Moskal:2000gj,
  Moskal:2000pu, Czerwinski:2014yot}, also give us a good ground to
study the $\eprime N$ interaction.

In this study, we theoretically investigate possibilities of observing
a signal of an $\eprime n$ bound state in the photoproduction cross
sections of $\eta$ and $\eprime$ mesons on a deuteron target, $\gamma
d \to \eta n p$ and $\eprime n p$, using the formulation developed in
Refs.~\cite{Jido:2009jf, Jido:2010rx, YamagataSekihara:2012yv,
  Jido:2012cy}.  For this purpose, we consider forward proton emission
so as to make a kinetically favored condition for the generation of
the $\eprime n$ bound state.  As for the production process, we take
into account a single-scattering $\eta ^{( \prime )}$ photoproduction
on a bound proton and double scatterings with the exchange of $\eta
^{( \prime )}$ meson, which is produced on a bound proton in the first
step.  We employ the linear sigma model~\cite{Sakai:2013nba,
  Sakai:2014zoa} so as to calculate the $\eprime N$ interaction and
its scattering amplitude.  Then we compare the signal of the $\eprime
n$ bound state in the $\gamma d \to \eta n p$ reaction to the
quasifree $\eprime$ contributions in the $\eprime n p$ reaction.

This paper is organized as follows.  In Sec.~\ref{sec:2} we develop
our formulation of the cross sections of the $\eta$ and $\eprime$
photoproductions on proton and deuteron targets.  The $\eprime N$
interaction in our effective model is also briefly introduced in this
section.  Next, in Sec.~\ref{sec:3} we show our results of the $\eta$
and $\eprime$ photoproduction cross sections on a deuteron target and
discuss possibilities of observing the signal of the $\eprime n$ bound
state by comparing the signal with the quasifree $\eprime$
contribution.  In this section we also discuss the behavior of our
results in several experimental conditions and model parameters.
Section~\ref{sec:4} is devoted to the conclusion of this study.

\section{Formulation}
\label{sec:2}

In this section we formulate the cross sections of the $\eta ^{(
  \prime )}$ photoproduction on the deuteron and proton targets.
First, we consider the deuteron target case and discuss the diagrams
for the photoproduction of the $\eprime n$ bound state off the
deuteron in Sec.~\ref{sec:2-1}.  Next, in Sec.~\ref{sec:2-2} we
explain our approach to calculate the $\eprime N$ scattering
amplitude, in which an $\eprime N$ bound state can appear as a
resonance pole with appropriate model parameters of the linear sigma
model.  Finally, we go to the $\gamma p \to \eta ^{( \prime )} p$
reaction in Sec.~\ref{sec:2-3}, where we take into account the $\eta
^{( \prime )} N \to \eta ^{( \prime )} N$ rescattering process with
the amplitude developed in Sec.~\ref{sec:2-2}, and we fix the
parameters for the $\gamma p \to \eta ^{( \prime )} p$ reaction so as
to reproduce the experimental data.

\subsection{The $\bm{\gamma d \to \eta n p}$ and
  $\bm{\eprime n p}$ reactions}
\label{sec:2-1}

Let us first consider the $\eta ^{( \prime )}$ meson photoproduction
on the deuteron target, $\gamma d \to X p$ with $X = m n = \eta n$ or
$\eprime n$.  The differential cross section of the reaction is
expressed as
\begin{equation}
\frac{d^{2} \sigma _{\gamma d \to X p}}{d M_{X} d \Omega _{p}} 
= \frac{p_{p} p_{m}^{\ast} M_{p} M_{n}}{4 \Eglab W_{3}} 
\frac{1}{( 2 \pi )^{5}} 
\int d \Omega _{n}^{\ast} | T_{\gamma d \to X p} |^{2} ,
\label{eq:dsdMdOmega}
\end{equation}
where $M_{X} = M_{m n} = M_{\eta n}$ or $M_{\eprime n}$ is the
invariant mass of $X$, $\Omega _{p}$ is the solid angle for the
momentum of the final-state proton in the global center-of-mass frame,
$M_{p}$ and $M_{n}$ are the proton and neutron masses, respectively,
$\Eglab$ is the photon energy in the laboratory frame, i.e., the
deuteron rest frame, $W_{3}$ is the total energy obtained as $W_{3} =
\sqrt{M_{d}^{2} + 2 M_{d} \Eglab }$ with the deuteron mass $M_{d}$,
$\Omega _{n}^{\ast}$ is the solid angle for the momentum of the
neutron in the $m$-$n$ center-of-mass frame, and $T_{\gamma d \to X
  p}$ is the scattering amplitude for the reaction $\gamma d \to X p$.
The magnitude of the momenta of the final-state proton $p_{p}$ and
the meson $p_{m}^{\ast}$ are evaluated in the global center-of-mass
frame and in the $m$-$n$ center-of-mass frame, respectively, and they
are expressed as
\begin{equation}
p_{p} = \frac{\lambda ^{1/2} ( W_{3}^{2} , \, M_{p}^{2}, \, M_{X}^{2})}{2 W_{3}} ,
\quad 
p_{m}^{\ast} 
= \frac{\lambda ^{1/2} ( M_{X}^{2} , \, M_{m}^{2}, \, M_{n}^{2})}{2 M_{X}} ,
\end{equation}
with the \Kaellen function $\lambda (x, \, y, \, z) = x^{2} + y^{2} +
z^{2} - 2 x y - 2 y z - 2 z x$ and the meson mass $M_{m}$.

\begin{figure}[t]
  \centering
  \PsfigII{0.135}{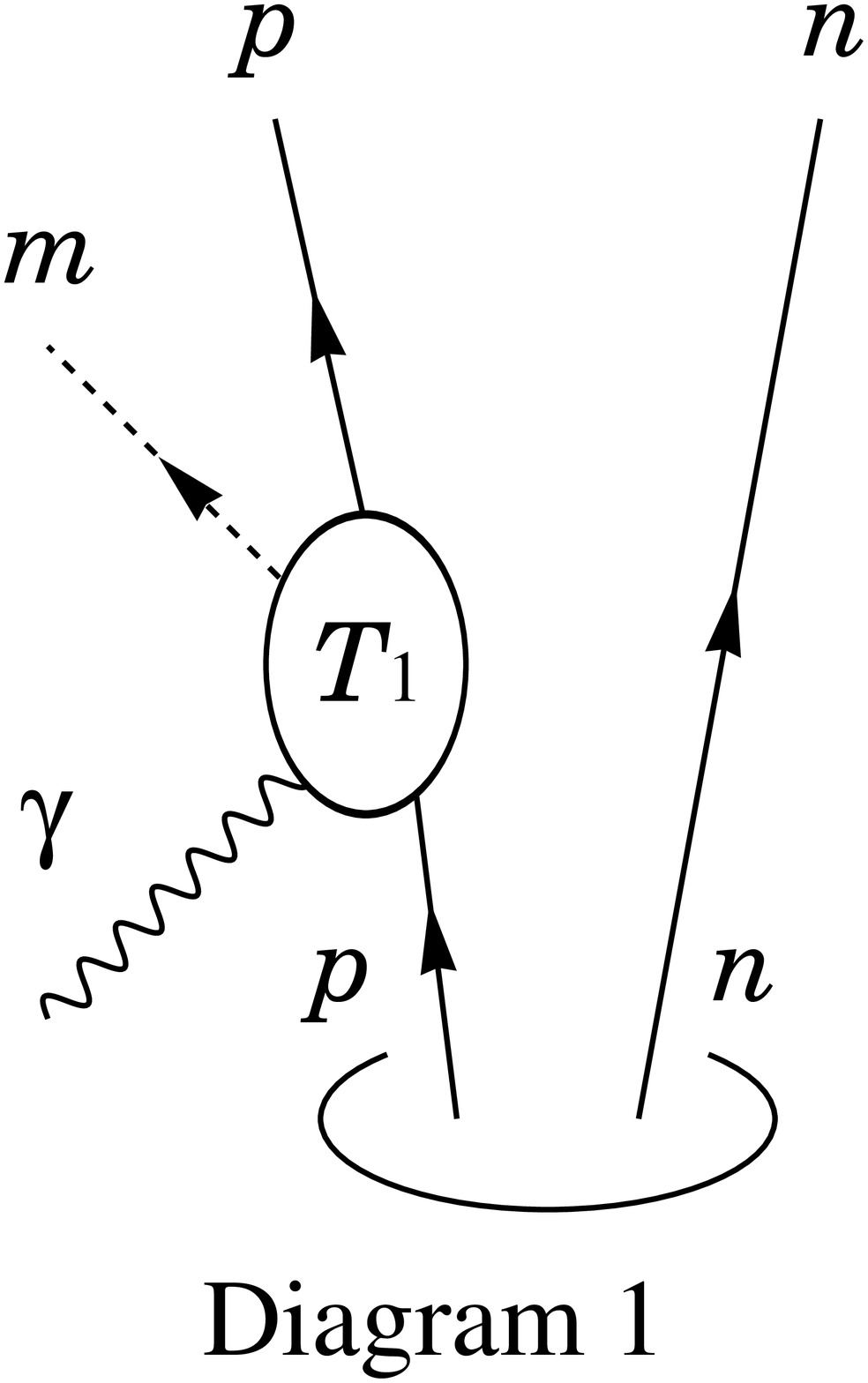} 
  \PsfigII{0.135}{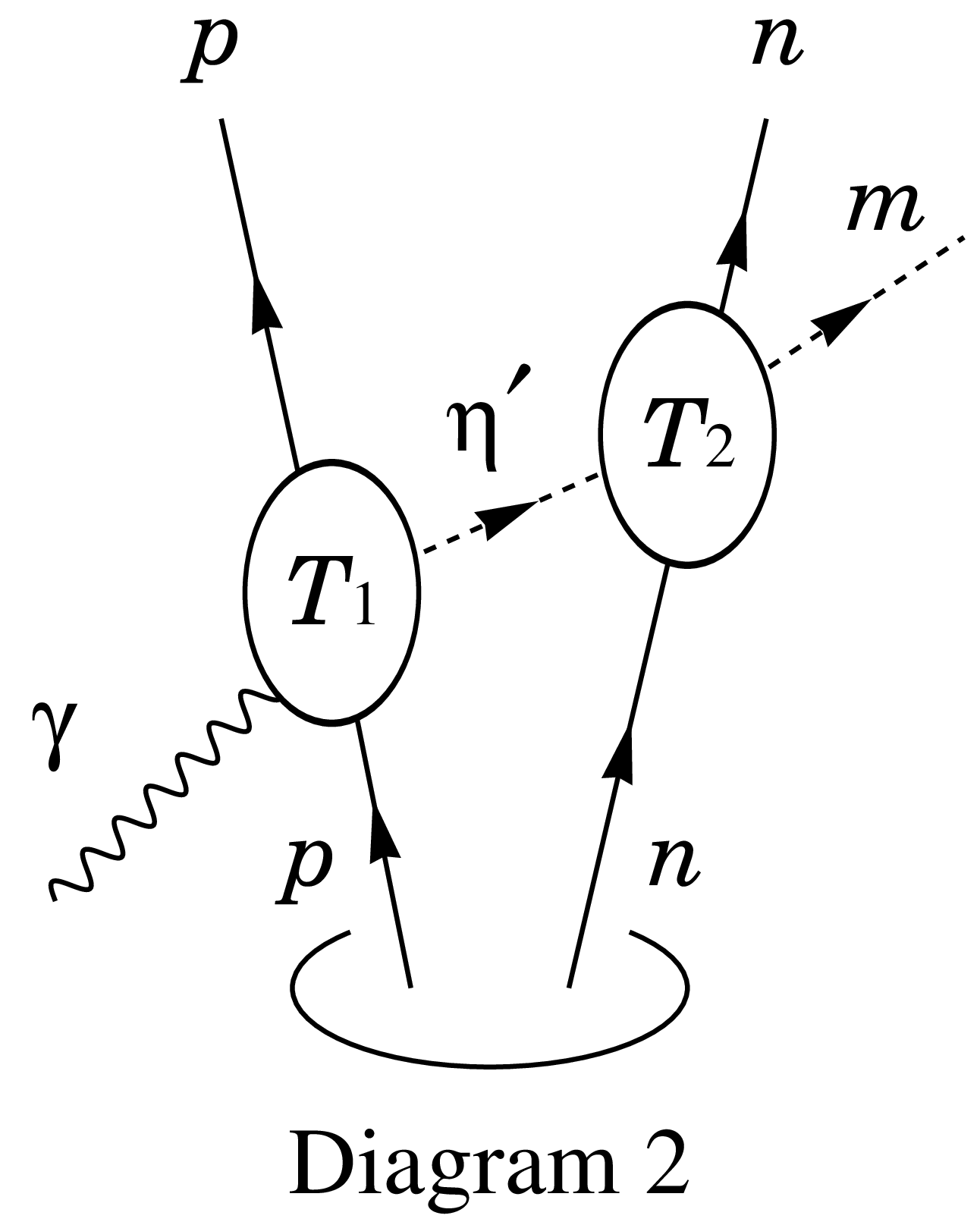} 
  \PsfigII{0.135}{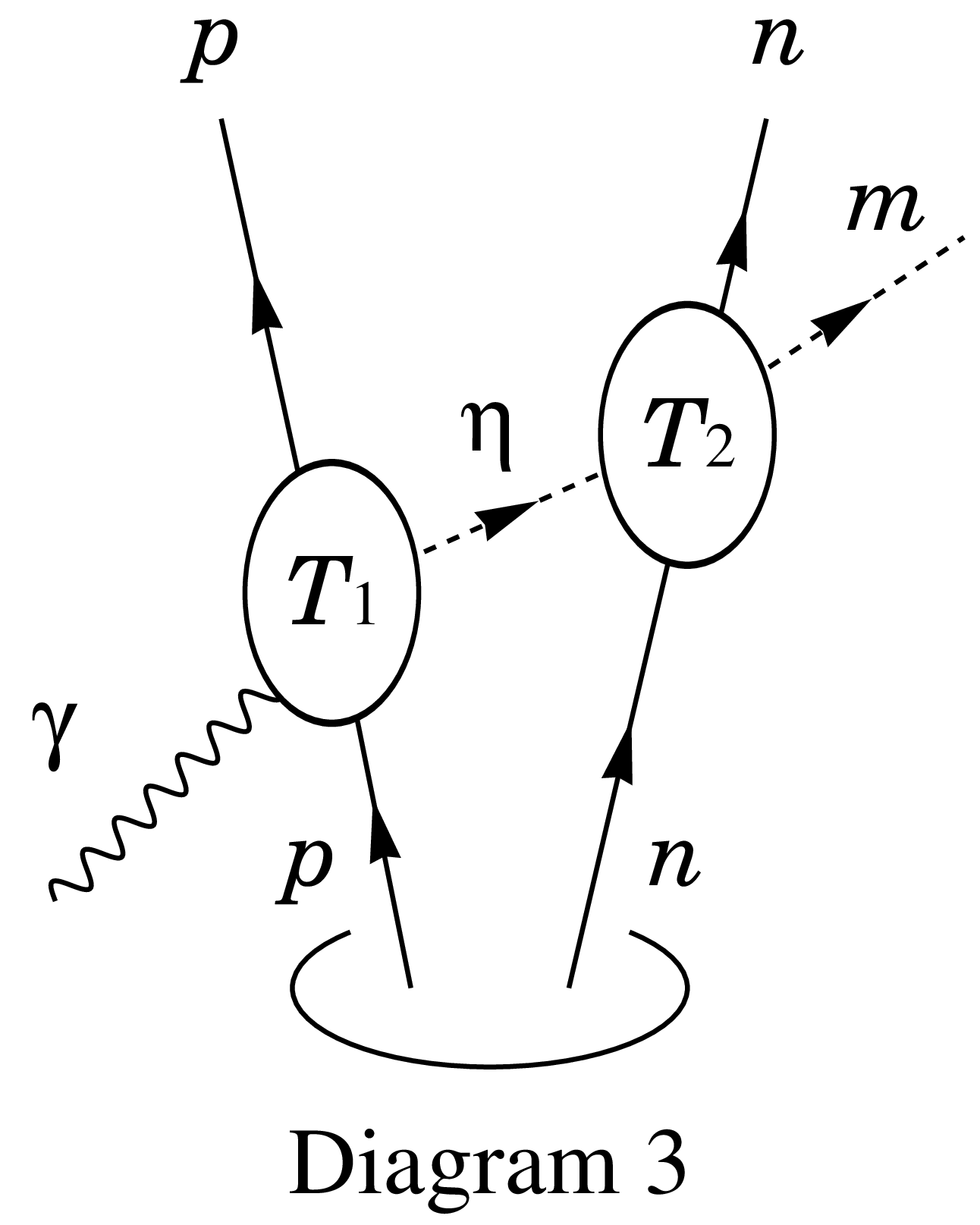} 
  \caption{Feynman diagrams for the $\gamma d \to m n p$ reaction with
    $m = \eta$ or $\eprime$.  Here $T_{1}$ and $T_{2}$ are the
    $\gamma p \to m p$ and $m n \to m n$ scattering amplitudes,
    respectively.}
  \label{fig:gamma-d}
\end{figure}

In this study we are interested in the photoproduction of an $\eta
^{\prime} n$ bound state with forward proton emission, so we calculate
the cross sections by considering kinetically favored amplitudes,
which are diagrammatically shown in Fig.~\ref{fig:gamma-d}.  Namely,
we take into account a single-scattering $\eta ^{( \prime )}$
photoproduction on a bound proton and double scatterings with the
exchange of $\eta ^{( \prime )}$ meson, which is produced on a bound
proton in the first step.  Since we require a fast proton in the
forward direction, we can safely neglect the final-state interaction
between proton and neutron.  In addition, as we will see later, the
$\eprime$ exchange is most important, since $\eprime$ in the
intermediate state goes almost on its mass shell at $M_{X} \approx
M_{\eprime} + M_{n}$.  On the other hand, the $\eta$ exchange is
suppressed due to its largely off-shell propagation.  This means that
exchanges of other mesons such as $\pi$ should be suppressed more. We
also note that we do not consider diagrams of $\eta$ and $\eta
^{\prime}$ photoproductions on a bound neutron.  This is because in
this condition the final-state neutron should go in forward direction
with large momentum while the final-state proton would be slow and its
scattering angle would not be restricted to forward due to the
kinematics, which can easily be suppressed by the experimental setup.

Thus, we calculate the scattering amplitude $T_{\gamma d \to X p}$ as
\begin{equation}
T_{\gamma d \to X p} = 
\mathcal{T}_{1}^{(m)} + \mathcal{T}_{2}^{(m)} + \mathcal{T}_{3}^{(m)} ,
\end{equation}
where the subscripts $1$, $2$, and $3$ corresponds to the number of
the diagrams in Fig.~\ref{fig:gamma-d}.  The expression of each
amplitude is obtained in a similar manner to that in
Refs.~\cite{Jido:2009jf, Jido:2010rx, YamagataSekihara:2012yv,
  Jido:2012cy}.

The first term $\mathcal{T}_{1}^{(m)}$, corresponding to the single
scattering, is evaluated as
\begin{equation}
  \mathcal{T}_{1}^{(m)} = T_{\gamma p \to m p} ( W_{2} )
  \times \deutwf ( \bm{p}_{n} ) , 
\end{equation}
with the $s$-wave $\gamma p \to m p$ scattering amplitude $T_{\gamma p
  \to m p}$ denoted by $T_1$ in Fig.~\ref{fig:gamma-d} and the
deuteron wave function in momentum space, $\deutwf (\bm{q} )$, which
is given in the deuteron rest frame.  Therefore, for the evaluation of
the deuteron wave function $\deutwf ( \bm{p}_{n} )$ we have to
calculate the neutron momentum in the final state, $\bm{p}_{n}$, in
the laboratory frame.  The energy $W_{2}$ is calculated as
\begin{equation}
  W_{2} = \sqrt{ \left ( p_{m}^{\mu} + p_{p}^{\mu} \right ) ^{2} } ,
\end{equation}
where $p_{m}^{\mu}$ and $p_{p}^{\mu}$ are the momenta of the meson $m$
and proton in the final state, which can be evaluated from the
final-state phase space.

The second term $\mathcal{T}_{2}^{(m)}$ corresponds to the double
scattering with the $\eprime$ exchange and evaluated as
\begin{align}
  \mathcal{T}_{2}^{(m)} = & T_{\gamma p \to \eprime p} ( W_{2}^{\prime} )
  T_{\eprime n \to m n} ( M_{X} )
  \notag \\
  & \times \int \frac{d^{3} q}{(2 \pi )^{3}} 
  \frac{\deutwf ( \bm{q} + \bm{p}_{p} - \bm{k} )}
       {q^{2} - M_{\eprime}^{2} + i \epsilon} ,
       \label{eq:T2}
\end{align}
with the $\eprime n \to m n$ amplitude $T_{\eprime n \to m n}$ denoted
by $T_2$ in Fig.~\ref{fig:gamma-d}, for which we employ an effective
model described in the next subsection, the photon and final-state
proton momenta in the laboratory frame $\bm{k} = ( 0, \, 0, \, \Eglab
)$ and $\bm{p}_{p}$, respectively, and an infinitesimal positive value
$\epsilon$.  The energy $W_{2}^{\prime}$ is approximated as
\begin{equation}
  W_{2}^{\prime} \approx \sqrt{M_{p}^{2} + 2 M_{p} \Eglab} ,
\end{equation}
by assuming that the initial-state bound proton is at rest on its mass
shell.  The energy carried by the exchanged meson, $q^{0}$, should be
fixed in appropriate models.  In this study, we employ two approaches.
The first one is the Watson approach~\cite{Watson:1953zz}, which gives
us~\cite{Jido:2012cy}
\begin{equation}
  q^{0} = M_{p} + \Eglab - p_{p}^{0} ,
  \label{eq:q0}
\end{equation}
in the laboratory frame.  In the second approach, we employ the
truncated Faddeev approach as done in Ref.~\cite{Miyagawa:2012xz}, in
which we have
\begin{equation}
q^{0} = M_{d} + E_{\gamma}^{\rm lab} - p_{p}^{0} 
- M_{n} - \frac{| \bm{q} + \bm{p}_{p} - \bm{k} |^{2}}{2 M_{n}} , 
\label{eq:q0_Faddeev}
\end{equation}
in the laboratory frame.  Here we refer to the former (latter)
treatment as option A (B).  The details are given in
Ref.~\cite{Jido:2012cy}.

The third term $\mathcal{T}_{3}^{(m)}$ corresponds to the double
scattering with the $\eta$ exchange and is evaluated as
\begin{align}
\mathcal{T}_{3}^{(m)} 
= & T_{\gamma p \to \eta p} ( W_{2}^{\prime} ) T_{\eta n \to m n} ( M_{X} )
\notag \\
& \times 
\int \frac{d^{3} q}{(2 \pi )^{3}} 
\frac{\deutwf ( \bm{q} + \bm{p}_{p} - \bm{k} )}
{q^{2} - M_{\eta}^{2} + i \epsilon} .
\label{eq:T3}
\end{align}
Here the energy carried by the exchanged meson, $q^{0}$, is fixed in
the same manner as in the second term, $\mathcal{T}_{2}^{(m)}$, with
the option A~\eqref{eq:q0} or B~\eqref{eq:q0_Faddeev}.

In our calculation, both $T_{\gamma p \to \eta p}$ and $T_{\gamma p
  \to \eprime p}$ can be factorized out of the integral because we
assume it not to depend on the internal energy nor scattering angle.
In a more realistic case, both $T_{\gamma p \to \eta p}$ and
$T_{\gamma p \to \eprime p}$ depend on them and thus should be in
principle inside the integral.  Nevertheless, the forward proton
emission of this reaction, i.e., the scattering angle of the
final-state proton, $\theta _{p}$, being around $0$ degree, indicates
that neglecting the angular dependence is enough good as a first-order
approximation.  On the other hand, the energy $W_{2}^{\prime}$ as a
parameter of the amplitudes $T_{\gamma p \to m p}$ can be fixed by
assuming that the initial-state bound proton in the first scattering
is at rest on its mass shell, as done in Ref.~\cite{Jido:2009jf}.

For the deuteron wave function, we neglect the $d$-wave component and
we use a parameterization of the $s$-wave component given by an
analytic function~\cite{Lacombe:1981eg} as
\begin{equation} 
\deutwf (\bm{q}) =
\sum _{j=1}^{11} \frac{C_{j}}{\bm{q}^{2} + m_{j}^{2}} ,
\label{eq:deutWF}
\end{equation}
with $C_{j}$ and $m_{j}$ determined in~\cite{Machleidt:2000ge}.

\subsection{The $\bm{\eprime N}$ scattering amplitude}
\label{sec:2-2}

Next we formulate the $\eprime N$ scattering amplitude around the
$\eprime N$ threshold. In this study we consider an $s$-wave $\eprime
N$-$\eta N$ coupled-channels problem, since the $\eta N$ channel can
be important to the $\eprime N$ scattering amplitude as the closest
open channel in $s$ wave.  In this study, we employ the $\eprime N$
amplitude obtained from the linear sigma model with unitarization
according to the approach developed in Refs.~\cite{Sakai:2013nba,
  Sakai:2014zoa}.  The scattering amplitude $T_{i j}$ is labeled by
the channel indices $i$ and $j$ as $i=1$ ($2$) for $\eprime N$ ($\eta
N$).  Here we note that we employ the physical masses for nucleons to
calculate quantities, so the nucleon mass $M_{N}$ is equal to $M_{p}$
for the $\eta ^{( \prime )} p$ reaction and to $M_{n}$ for the $\eta
^{( \prime )} n$ reaction in the following formulation, while the
interaction term is constructed with isospin symmetry.

According to Refs.~\cite{Sakai:2013nba, Sakai:2014zoa}, we construct
an interaction kernel from the linear sigma model as
\begin{equation}
  V_{1 1} = - \frac{6 g B}{\sqrt{3} m_{\sigma 0}^{2}} ,
  \quad 
  V_{1 2} = V_{2 1} = + \frac{6 g B}{\sqrt{6} m_{\sigma 8}^{2}} ,
  \quad 
  V_{2 2} = 0 ,
  \label{eq:Vij}
\end{equation}
where constants $g$, $B$, $m_{\sigma 0}$, and $m_{\sigma 8}$ determine
the strength of the interaction; $g$ is the coupling constant for the
$\sigma N N$ vertex, $B$ represents the contribution from the
$\text{U}_{\rm A} ( 1 )$ anomaly, and $m_{\sigma 0}$ and $m_{\sigma
  8}$ are the masses of the singlet and octet sigma mesons exchanged
between $\eta ^{( \prime )}$ and $N$.  These parameters are fixed as
$g = 7.67$, $B = 0.984 \gev$, $m_{\sigma 0} = 0.7 \gev$, and
$m_{\sigma 8} = 1.23 \gev$~\cite{Sakai:2013nba}.

Here, we note that the contribution from the $\eta N$ channel is not
so large because the mixing angle between the $\eta$ and $\eta'$ is
small and the transition of the $\eta' N$ into $\eta N $ governed by
Eq.~(\ref{eq:Vij}) is suppressed by the large mass of the octet scalar
meson $m_{\sigma 8}$.  This means that the following result would not
depend so much on the details of the treatment of the $\eta N$
channel.

We use this tree-level interaction as an interaction kernel, and solve
the scattering equation to obtain the scattering amplitude $T_{i j}
(w)$:
\begin{equation}
T_{i j} ( w ) 
= V_{i j} + \sum _{k = 1}^{2} V_{i k} G_{k} ( w ) T_{k j} ( w ) , 
\label{eq:BSeq}
\end{equation}
where $w$ is the center-of-mass energy and $G_{i}$ is the $\eta ^{(
  \prime )} N$ loop function.  It is important that the tree-level
amplitude $V_{i j}$ is independent of the external momentum [see
Eq.~\eqref{eq:Vij}], and thus the scattering equation becomes an
algebraic equation.  For the loop function $G_{i}$, we employ a
covariant expression as
\begin{equation}
  G_{i} ( w ) \equiv 
  i \int \frac{d^{4} q}{(2 \pi )^{4}} 
  \frac{2 M_{N}} {[(P - q)^{2} - M_{N}^{2}] (q^{2} - M_{i}^{2})} ,
\end{equation}
with $P^{\mu} = (w , \, \bm{0})$, $M_{1} = M_{\eprime}$, and $M_{2} =
M_{\eta}$, and the loop function is calculated with the dimensional
regularization as
\begin{align}
& G_{i} (w) = \frac{2 M_{N}}{16 \pi ^{2}} \bigg [ a_{i} ( \mu _{\rm reg} ) 
+ \ln \left ( \frac{M_{N}^{2}}{\mu _{\rm reg}^{2}} \right )  
\notag \\ &
+ \frac{w^{2} + M_{i}^{2} - M_{N}^{2}}{2 w^{2}} 
\ln \left ( \frac{M_{i}^{2}}{M_{N}^{2}} \right ) 
\notag \\
& - \frac{\lambda ^{1/2} (w^{2}, \, M_{N}^{2}, \, M_{i}^{2})}{w^{2}} 
\text{artanh} 
\left ( \frac{\lambda ^{1/2} (w^{2}, \, M_{N}^{2}, \, M_{i}^{2})}
{M_{N}^{2} + M_{i}^{2} - w^{2}} \right ) 
\bigg ] ,
\label{eq:Gdim_explicit}
\end{align}
with the subtraction constant $a_{i}$ at the regularization scale $\mu
_{\rm reg}$, which is set as $\mu _{\rm reg} = M_N$.  In this study
they are fixed by the natural renormalization scheme developed in
Ref.~\cite{Hyodo:2008xr} so as to exclude the Castillejo-Dalitz-Dyson
(CDD) pole contribution from the loop function.  This can be achieved
by requiring $G_{i} ( w = M_{N} ) = 0$ for every channel $i$.

In this construction, a sufficient attraction between $\eprime$ and
$N$ leads to an $\eprime N$ bound state described by a pole of the
scattering amplitude $T_{i j}(w)$ with its residue $g_{i} g_{j}$:
\begin{equation}
T_{i j} ( w ) = \frac{g_{i} g_{j}}{w - w_{\rm pole}}
+ ( \text{regular at }w = w_{\rm pole} ) . 
\label{eq:amp_pole}
\end{equation}
The residue $g_{i}$ can be interpreted as the coupling constant of the
$\eprime N$ bound state to the $i$ channel.  The coupling constant
$g_{i}$ is further translated into the so-called compositeness $X_{i}$
via the two-body wave function so as to measure the fraction of the
two-body component~\cite{Hyodo:2011qc, Aceti:2012dd, Hyodo:2013nka,
  Sekihara:2014kya, Sekihara:2015gvw}.  Namely, in the present
formulation the two-body wave function in channel $i$ in momentum
space $ \tilde{\Psi}_{i}( \bm{q} )$ is proportional to the coupling
constant $g_{i}$~\cite{Gamermann:2009uq, YamagataSekihara:2010pj} as
\begin{equation}
  \tilde{\Psi}_{i} ( \bm{q} ) 
  = \frac{g_{i} \sqrt{4 M_{N} w_{\rm pole}}}{w_{\rm pole}^{2}
    - [ \omega _{i} ( \bm{q} ) + \Omega _{N} ( \bm{q} ) ]^{2} } .
\end{equation}
Then, the compositeness is defined as the norm of $\tilde{\Psi}_{i} (
\bm{q} )$, and its expression is
\begin{align}
  X_{i} = & \int \frac{d^{3} q}{( 2 \pi )^{3}}
  \frac{\omega _{i} ( \bm{q} ) + \Omega _{N} ( \bm{q} )}
       {2 \omega _{i} ( \bm{q} ) \Omega _{N} ( \bm{q} )}
  \left [ \tilde{\Psi}_{i} ( \bm{q} ) \right ]^{2}
  \notag \\
  = & - g_{i}^{2} \frac{d G_{i}}{d w} ( w = w_{\rm pole} ) ,
\end{align}
where $\omega _{i} ( \bm{q} ) \equiv \sqrt{M_{i}^{2} + \bm{q}^{2}}$
and $\Omega _{N} ( \bm{q} ) \equiv \sqrt{M_{N}^{2} + \bm{q}^{2}}$.
Here we note that the compositeness as well as the wave function is a
scheme dependent quantity, i.e., we can uniquely determine it when we
fix the model space, interaction, and loop function.  Since we take
into account only the $\eprime N$ and $\eta N$ channels in the present
model, the sum of the norms for the $\eprime N$ and $\eta N$ channels,
$X_{1} + X_{2}$, coincides with the normalization of the total
bound-state wave function $| \Psi \rangle$ as
\begin{equation}
  \langle \Psi ^{\ast} | \Psi \rangle
  = X_{1} + X_{2} = 1 .
  \label{eq:sum_rule}
\end{equation}
In this sense, one can deduce the structure by comparing the value of
the compositeness with unity.  Besides, we may take into account
missing channels, which do not appear as explicit degrees of freedom
in the model space, by employing an energy dependent two-body
interaction, as such a missing channel inevitably brings energy
dependence to the two-body interaction~\cite{Sekihara:2014kya,
  Hyodo:2008xr}.

\begin{table}
  \caption{Pole position $w_{\rm pole}$, coupling constant $g_{i}$ ($i
    = \eprime N$, $\eta N$), and compositeness $X_{i}$ of the $\eprime
    N$ bound state in the present model.  }
  \label{tab:1}
  \begin{ruledtabular}
    \begin{tabular*}{8.6cm}{@{\extracolsep{\fill}}lc}
      $w_{\rm pole}$ [MeV] & $1889.5 - 6.3 i$ \\
      $g_{\eprime N}$ & $2.40 + 0.45 i$ \\
      $g_{\eta N}$ & $-0.54 - 0.07 i \phantom{-}$ \\
      $X_{\eprime N}$ & $1.01 + 0.00 i$ \\
      $X_{\eta N}$ & $-0.01 + 0.00 i \phantom{-}$ \\
    \end{tabular*}
  \end{ruledtabular}
\end{table}

The values of the pole position, coupling constant, and compositeness
of the $\eprime N$ bound state in the present model are listed in
Table~\ref{tab:1}.  As one can see, the pole position $w_{\rm pole}$
has a small imaginary part as a decay of the $\eprime N$ bound state
to the $\eta N$ channel, and the value is consistent with the
experimental implication in Ref.~\cite{Moyssides:1983}.  The modulus
of the $\eprime N$ coupling constant is about five times larger than
that of the $\eta N$ one.  Since the $\eprime N$ compositeness
$X_{\eprime N}$ is close to unity with a negligible imaginary part,
the $\eprime N$ bound state in the present model parameter is indeed
dominated by the $\eprime N$ component.

\subsection{The $\bm{\gamma p \to \eta p}$ and
  $\bm{\eprime p}$ scattering amplitudes}
\label{sec:2-3}

\begin{figure}[!b]
  \centering
  \PsfigII{0.19}{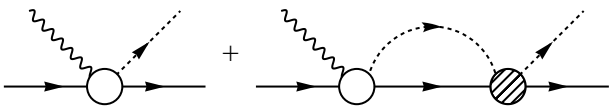} 
  \caption{Feynman diagrams for the $\gamma p \to m p$ reaction with
    $m = \eta$ or $\eprime$.  Here the solid, dashed, and wavy lines
    represent the proton, $m$, and photon, respectively.  The open and
    shaded circles correspond to the $\gamma p \to m p$ and $m p \to m
    p$ amplitudes, respectively.}
  \label{fig:Tg}
\end{figure}

Finally, let us consider photoproductions of $\eta$ and $\eprime$ on a
proton target.  In this study, we introduce the rescattering of $\eta
^{( \prime )} p$ in the final state of the $\gamma p \to m p$ reaction
with $m = \eta$ or $\eprime$, as done in Ref.~\cite{Nacher:1998mi}.
Namely, with the $\eta ^{( \prime )} N \to \eta ^{( \prime )} N$
amplitude developed in the previous subsection, we construct the
scattering amplitude $T_{\gamma p\to m p}$ in the approach
diagrammatically shown in Fig.~\ref{fig:Tg}, which is expressed as
\begin{equation}
  T_{\gamma p \to i} ( W ) = V_{\gamma i} + \sum _{j = 1}^{2} V_{\gamma j}
  G_{j} ( W ) T_{j i} ( W ) .
  \label{eq:Tg}
\end{equation}
Here $W$ is the center-of-mass energy, $T_{j i}$ and $G_{j}$ are the
$\eta ^{( \prime )} p \to \eta ^{( \prime )} p$ scattering amplitude
and loop function developed in the previous subsection, respectively,
and the channel index $i = 1$ ($2$) indicates the $\eprime p$ ($\eta
p$) channel.  In general we may take different subtraction constants
for the loop functions $G_i$ in Eqs.~(\ref{eq:BSeq}) and
(\ref{eq:Tg}), but the same subtraction constant is used in this
study.  In contrast, the $\gamma p \to i$ part $V_{\gamma i}$ is
unknown model parameter.

\begin{figure}[!t]
  \centering
  \Psfig{8.6cm}{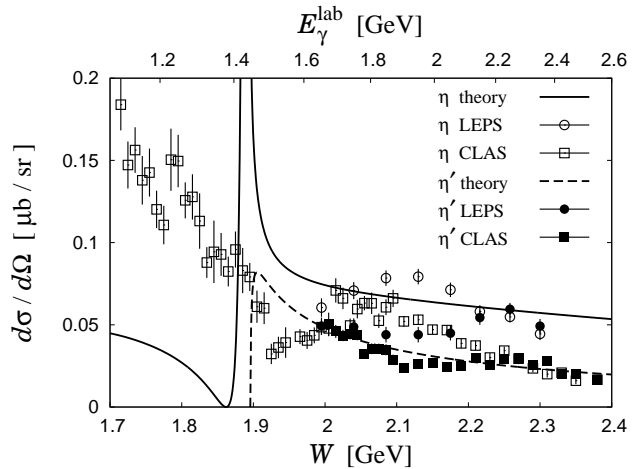} 
  \caption{Differential cross sections of $\gamma p \to \eta p$ and
    $\eprime p$ reactions calculated with the amplitudes in
    Eq.~\eqref{eq:Tg} and comparison with the experimental data in
    Refs.~\cite{Williams:2009yj, Sumihama:2009gf}.  The experimental
    data are taken with the scattering angle $- 0.8 < \cos \theta
    _{m}^{\rm cm} < -0.7$ for both $\eta$ and $\eprime$
    photoproductions with $\theta _{m}^{\rm cm}$ being the meson angle
    in the center-of-mass frame. }
  \label{fig:data}
\end{figure}

In this study we fix $V_{\gamma p \to i}$ by using the experimental
data of the differential cross section for the reaction $\gamma p \to
m p$, which is expressed as
\begin{equation}
  \frac{d \sigma _{\gamma p \to m p}}{d \Omega} 
  = \frac{p_{\rm cm}^{\prime} M_{p}}{16 \pi ^{2} \Eglab W}
  | T_{\gamma p \to m p} |^{2} .
\label{eq:dsdO_p}
\end{equation}
Here $\Omega$ is the solid angle for the momentum of the final-state
proton in the center-of-mass frame and the total energy $W$ is
obtained as $W = \sqrt{ M_{p}^{2} + 2 M_{p} \Eglab }$.  The magnitude
of the momentum of the final-state proton in the center-of-mass frame,
$p_{\rm cm}^{\prime}$, can be calculated as
\begin{equation}
p_{\rm cm}^{\prime} 
= \frac{\lambda ^{1/2} ( W^{2}, \, M_{p}^{2}, \, M_{m}^{2})}{2 W} .
\end{equation}
Here we note that, since we mainly concentrate on the forward proton
emission, we may need only the scattering amplitude at a certain
angle.  Furthermore, in this study we are interested in the ratio of
the $\eprime n$ bound state signal to the $\eprime$ quasifree
contribution.  In this sense, regarding the $\gamma p \to i$ part
$V_{\gamma i}$ to be constant is enough for our purpose to calculate
the relative strength between the $\eprime n$ bound state signal and
the $\eprime$ quasifree contribution in the forward proton emission.
Thus, we fix two fitting parameters $V_{\gamma 1}$ and $V_{\gamma 2}$
so as to reproduce the experimental data.  For the forward proton
emission, we use the experimental data on the $\gamma p \to \eprime p$
and $\eta p$ reactions in the scattering angle $-0.8 < \cos \theta
_{m}^{\rm cm} < -0.7$ with the $\eta ^{( \prime )}$ scattering angle
$\theta _{m}^{\rm cm}$~\cite{Williams:2009yj, Sumihama:2009gf}.  As we
will see in the numerical results, the $\gamma p \to \eprime p$
reaction is most important, so we give more weight to the data of the
$\gamma p \to \eprime p$ reaction.  From the fit with the parameters
in the $\eta ^{( \prime )} N \to \eta ^{( \prime )} N$ amplitude ($g =
7.67$, $B = 0.984 \gev$, $m_{\sigma 0} = 0.7 \gev$, and $m_{\sigma 8}
= 1.23 \gev$), we take the following parameters:
\begin{equation}
  V_{\gamma 1} = 0.348 \gev ^{-1} ,
  \quad 
  V_{\gamma 2} = 0.354 \gev ^{-1} ,
\end{equation}
which reproduce the experimental cross sections with forward proton
emission above the $\eprime p$ threshold in
Ref.~\cite{Williams:2009yj, Sumihama:2009gf}, as shown in
Fig.~\ref{fig:data}.  We note that in Fig.~\ref{fig:data} we have a
prominent peak in the $\gamma p \to \eta p$ cross section below $W =
1.9 \gev$ corresponding to the signal of the $\eprime p$ bound state.
In actual experimental observation, this contribution should interfere
with others coming from the nonresonant background.  This may provide
a peak structure or a dip, generally a Fano resonance, depending on
the interference.\footnote{Actually, an enhancement of the
  differential cross section of the $\gamma p \to \eta p$ reaction was
  observed just below the $\eprime p$ threshold in
  experiments~\cite{Dugger:2002ft, Kashevarov:2016owq}, which was
  claimed to be attributable to an $S_{11}$ resonance in their
  analyses.  This might imply the signal of the $\eprime p$ bound
  state.}

We emphasize again that this strategy is sufficient for our purpose to
estimate the production ratio of the $\eprime n$ bound state compared
to the $\eprime$ quasifree contributions with forward proton emission.
Actually, around the $\eprime n$ threshold the strength of both the
bound state signal and the quasifree contribution is similarly
suppressed as the scattering angle increases, and hence a large
cancellation will take place when we take the signal to quasifree
ratio.

\section{Numerical Results}
\label{sec:3}

Now we calculate the differential cross section~\eqref{eq:dsdMdOmega}
for the $\gamma d \to m n p$ reaction with $m = \eta$ or $\eta
^{\prime}$.  We first perform theoretical studies of the signal for
the $\eprime n$ bound state in the photoproduction process in
Sec.~\ref{sec:3-1}.  In this section, after examining two options,
i.e., the Watson approach~\eqref{eq:q0} and the truncated Faddeev
approach~\eqref{eq:q0_Faddeev}, we investigate each diagram
contribution to the cross sections of the two reactions.  In addition,
we study how the signal of the $\eprime n$ bound state depends on the
strength of the $\eprime N$ interaction.  Then, in Sec.~\ref{sec:3-2}
we discuss how the signal of the $\eprime n$ bound state can be seen
in several experimental conditions.  We here show the dependence of
our results with respect to the initial photon energy and the
scattering angle of the final-state proton, and integrate the
differential cross section with respect to the scattering angle for
the forward proton emission.

Throughout this section, the initial photon energy $\Eglab$ and proton
scattering angle in the global center-of-mass frame $\theta _{p}^{\rm
  cm}$ are fixed as $\Eglab = 2.1 \gev$ and $\theta _{p}^{\rm cm} = 0$
degree, respectively, unless explicitly mentioned.

\subsection{Theoretical study of the $\bm{\eprime n}$ signal}
\label{sec:3-1}

\subsubsection{Signal of the $\eprime n$ bound state in two options}

\begin{figure}[!t]
  \centering
  \Psfig{8.6cm}{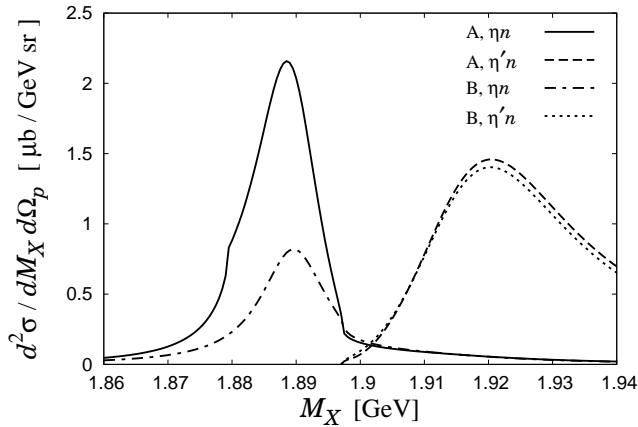} 
  \caption{Invariant mass spectra for the $\gamma d \to X p$ reactions
    with $X = \eta n$ and $\eprime n$.  The exchanged meson energy
    $q^{0}$ is fixed as in Eq.~\eqref{eq:q0}
    [Eq.~\eqref{eq:q0_Faddeev}] for the option A (B).  The initial
    photon energy is fixed as $\Eglab = 2.1 \gev$ and the proton
    scattering angle in the global center-of-mass is $\theta _{p}^{\rm
      cm}=0$ degree.  }
  \label{fig:AB}
\end{figure}

First of all, we examine two options of the exchanged meson energy: A
for the Watson approach~\eqref{eq:q0} and B for the truncated Faddeev
approach~\eqref{eq:q0_Faddeev}.  We calculate the differential cross
sections for the $\gamma d \to \eta n p$ and $\eprime n p$ reactions
in both two approaches as functions of the invariant mass $M_{X} =
M_{\eta n}$ and $M_{\eprime n}$, and the result is shown in
Fig.~\ref{fig:AB} in the range [$1.86 \gev$, $1.94 \gev$].  As one can
see from the figure, in both options A and B, we can clearly observe
the signal of the $\eprime n$ bound state in the $\eta n$ mass
spectrum below the $\eprime n$ threshold $\approx 1.897 \gev$, which
is comparable to the quasifree $\eprime$ contribution in the $\eprime
n$ mass spectrum above the threshold.  However, the strength of the
bound state signal is different in two options, while very similar
quasifree $\eprime$ contributions are found.  Namely, the option A (B)
gives a larger (smaller) signal of the $\eprime n$ bound state.  This
difference could be interpreted as a theoretical ambiguity in
calculating the differential cross section of the $\gamma d$ reaction
in the present formulation.

Here we should mention that in the option A we have a small cusp in
the $\eta n$ spectrum around $1.88 \gev$, which is an artificial
threshold in the Watson approach~\cite{Miyagawa:2012xz, Jido:2012cy}.
Since we are interested in the signal of the $\eprime n$ bound state
in clearer conditions, we employ only the option B, which gives
smaller signal of the bound state, in the following calculations.

Let us now numerically compare the contributions from the bound state
signal and from others above the $\eprime n$ threshold in option B.
This can be achieved by integrating the differential cross section in
appropriate ranges of the invariant mass $M_{X}$.  On the one hand,
the signal contribution is obtained by integrating $d^{2} \sigma
_{\gamma d \to \eta n p} / d M_{X} d \Omega _{p}$ in the range $[1.86
  \gev , \, M_{\eprime} + M_{n}]$, which results in $0.011 \mcb /
\text{sr}$.  On the other hand, the other contributions above the
$\eprime n$ threshold contains the quasifree $\eprime$ in the $\eprime
n$ spectrum and the tail of the $\eprime n$ bound state signal in the
$\eta n$ spectrum.  Thus, we integrate the sum of the cross section in
the two reactions, $d^{2} \sigma _{\gamma d \to \eta n p} / d M_{X} d
\Omega _{p} + d^{2} \sigma _{\gamma d \to \eprime n p} / d M_{X} d
\Omega _{p}$, in the range $[ M_{\eprime} + M_{n}, \, 2.0 \gev]$,
which results in $0.055 \mcb / \text{sr}$.  Therefore, we obtain the
ratio of the signal to other contributions as $0.011 / 0.055 = 0.20$.

\subsubsection{Contribution from each diagram}

\begin{figure}[!t]
  \centering
  \Psfig{8.6cm}{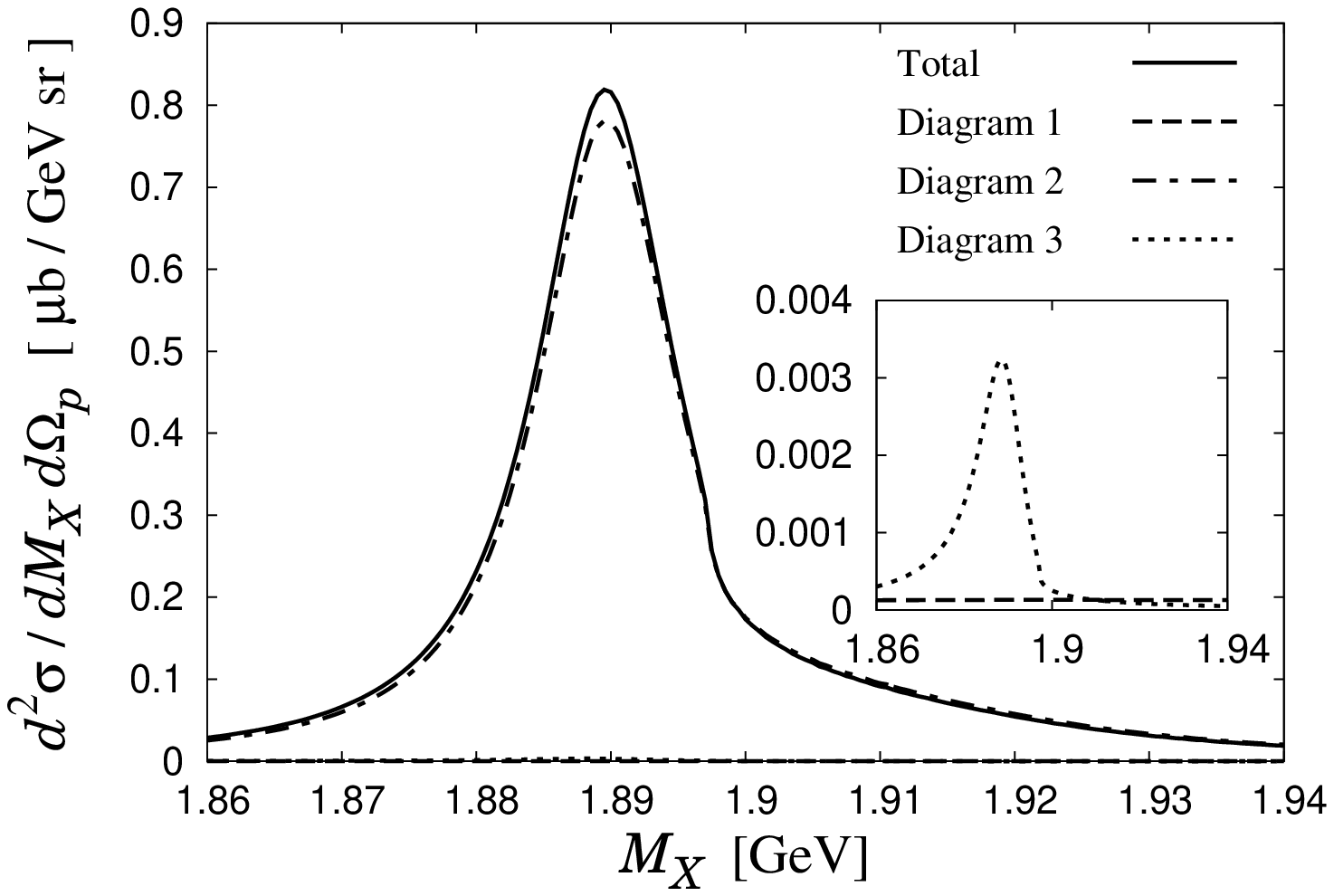} 
  \caption{Invariant mass spectrum for the $\gamma d \to \eta n p$
    reaction.  The initial photon energy is fixed as $\Eglab = 2.1
    \gev$ and the proton scattering angle in the global center-of-mass
    is $\theta _{p}^{\rm cm}=0$ degree.  The inset represents an
    enlarged figure.}
  \label{fig:eta}
\end{figure}

Next, we show in Fig.~\ref{fig:eta} the numerical result of each
diagram contribution to the differential cross section for the $\gamma
d \to \eta n p$ reaction~\eqref{eq:dsdMdOmega} as a function of the
invariant mass $M_{X} = M_{\eta n}$.  As one can see, we observe that
in this invariant mass region the cross section is dominated by the
diagram 2 in Fig.~\ref{fig:gamma-d}, i.e., the $\eta ^{\prime}$
exchange contribution.  This is because the invariant mass in this
region contains the $\eprime n$ threshold and thus the exchanged
$\eprime$ can go almost on its mass shell to generate an $\eprime n$
bound state.  On the other hand, both the diagrams 1 and 3 in
Fig~\ref{fig:gamma-d} are negligible.  The contribution from the
single scattering (diagram 1) is strongly suppressed by the deuteron
wave function.  Namely, in order to make the $\eta n$ invariant mass
as large as the $\eprime n$ threshold energy with the forward proton
emission only by the single scattering, we need anomalously large
Fermi motion of a bound neutron in the forward direction.  The $\eta$
exchange as the diagram 3 is also small because the exchanged $\eta$
cannot approach on its mass shell in the $\eprime n$ bound region with
forward proton emission and the magnitude of the $\eta n \to \eta n$
amplitude is small compared to that of the $\eprime n \to \eta n$ one
employed in the diagram 2.

\begin{figure}[!t]
  \centering
  \Psfig{8.6cm}{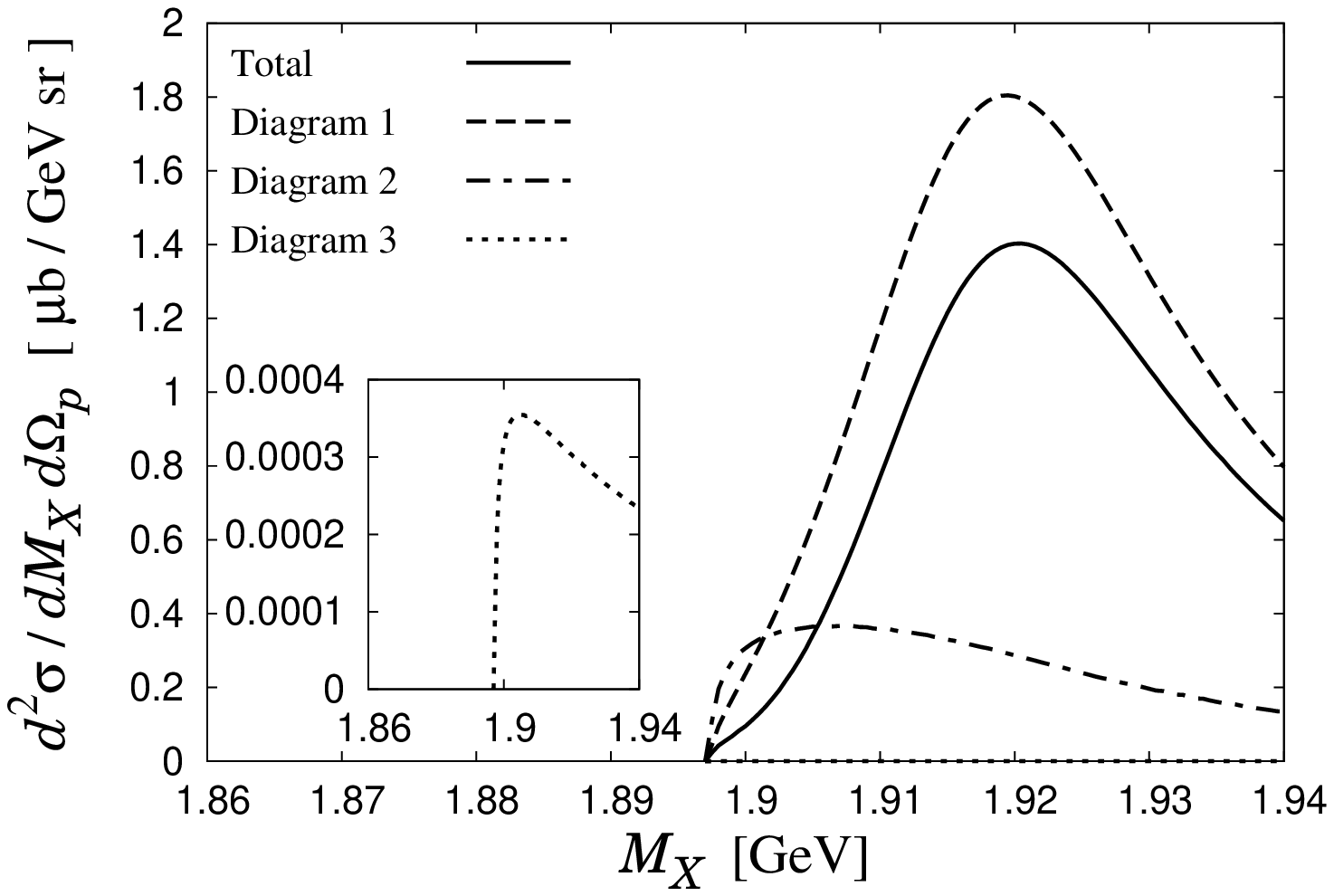} 
  \caption{Invariant mass spectrum for the $\gamma d \to \eprime n p$
    reaction.  The initial photon energy is fixed as $\Eglab = 2.1
    \gev$ and the proton scattering angle in the global center-of-mass
    is $\theta _{p}^{\rm cm}=0$ degree.  The inset represents an
    enlarged figure.}
  \label{fig:eta_prime}
\end{figure}

In Fig.~\ref{fig:eta_prime}, we show the numerical result of the
differential cross section for the $\gamma d \to \eprime n p$ reaction
around the $\eprime n$ threshold.  The cross section starts at the
$\eprime n$ threshold.  From the figure, we find that the quasifree
$\eprime$ contribution in the single scattering (diagram 1 in
Fig.~\ref{fig:gamma-d}) dominates the cross section.  This is caused
by the deuteron wave function; since a bound proton and a bound
neutron are almost at rest inside a deuteron, the $\eprime$ meson
produced by the $\gamma p^{\ast} \to \eprime p$ reaction with a bound
proton $p^{\ast}$ should be slow if the final-state proton goes the
forward angle with $\theta _{p}^{\rm cm} = 0$ degree, which makes the
invariant mass $M_{X}$ to be close to the $\eprime n$ threshold.
Besides, the tail of an $\eprime n$ bound state peak can make the
$\eprime$ exchange diagram (diagram 2) be a nonnegligible contribution
to the cross section as the dashed-dotted line in
Fig.~\ref{fig:eta_prime}.  On the other hand, the $\eta$ exchange
diagram negligibly contribute to the cross section due to a similar
reason as in the $\gamma d \to \eta n p$ reaction.

An interesting point is that we can observe the destructive
interference between the quasifree $\eprime$ photoproduction of the
single scattering and the $\eprime$ exchange contribution.  This means
the absorption of $\eprime$ produced on a bound proton into the bound
neutron inside the same deuteron.  Actually, we can easily find that
double scattering amplitude constructed with the imaginary part of the
$\eprime n \to \eprime n$ amplitude and the on-shell $\eprime$
exchange has opposite sign compared to the single scattering one.  The
present result provides us with an expectation that one may extract
information on the $\eprime N$ interaction from the quasifree
$\eprime$ production yield on a deuteron target compared to that on a
proton target.  We also expect large medium effects for $\eprime$ such
as the transparency ratio even in light nuclei.

\subsubsection{Dependence on the strength of the $\eprime N$ interaction}

\begin{table}
  \caption{Properties of the $\eprime n$ bound state with several
    values of the parameter $g$ or $m_{\sigma 8}$.  When changing the
    value of the parameter $g$ or $m_{\sigma 8}$, other parameters
    remain fixed.  We also consider the case that we introduce the
    contribution from the $\pi N$ channel~\cite{Sakai:2016}.  the The
    binding energy $B_{\rm E}$ and width $\Gamma$ are defined as
    $B_{\rm E} \equiv M_{\eprime} + M_{n} - \text{Re} \, w_{\rm pole}$
    and $\Gamma \equiv - 2 \, \text{Im} \, w_{\rm pole}$,
    respectively. }
  \label{tab:2}
  \begin{ruledtabular}
    \begin{tabular*}{8.6cm}{@{\extracolsep{\fill}}lccc}
      \multicolumn{4}{c}{Shift parameter $g$} \\
      $g$ & $g_{\eprime n}$ & $B_{\rm E}$ [MeV] & $\Gamma$ [MeV]
      \\
      \hline
      $5.0$ & \multicolumn{3}{l}{No structure}
      \\
      $6.0$ & \multicolumn{3}{l}{Cusp only}
      \\
      $7.0$ & $1.63 + 0.56 i$  &  $0.9$  &  $5.4$
      \\      
      $8.0$ & $2.71 + 0.43 i$  &  $12.8\phantom{0}$  &  $16.0\phantom{0}$
      \\
      $9.0$ & $3.49 + 0.40 i$  &  $31.8\phantom{0}$  &  $26.0\phantom{0}$
      \\
      \\
      \multicolumn{4}{c}{Shift parameter $m_{\sigma 8}$} \\
      $m_{\sigma 8}$ [GeV] & $g_{\eprime n}$ & $B_{\rm E}$ [MeV] & $\Gamma$ [MeV]
      \\
      \hline
      $0.9$ & $3.19 + 1.25 i$  &  $9.5$  &  $60.9\phantom{0}$
      \\
      $1.0$ & $2.79 + 0.91 i$  &  $8.8$  &  $34.4\phantom{0}$
      \\
      $1.1$ & $2.57 + 0.67 i$  &  $8.4$  &  $21.2\phantom{0}$
      \\
      $1.2$ & $2.43 + 0.49 i$  &  $8.0$  &  $14.1\phantom{0}$
      \\
      $1.3$ & $2.34 + 0.37 i$  &  $7.7$  &  $9.8$
      \\
      \\
      \multicolumn{4}{c}{Introduce $\pi N$ channel} \\
      & $g_{\eprime n}$ & $B_{\rm E}$ [MeV] & $\Gamma$ [MeV]
      \\
      \hline
      & $4.10+0.15i$ & $57.0$ & $14.5$ 
      \\
    \end{tabular*}
  \end{ruledtabular}
\end{table}

Now we see the dependence on the strength of the $\eprime n$
interaction for the peak structure of the $\eprime n$ bound state in
the $\gamma d \to p X$ reaction with $X = \eta n$ and $\eprime n$.
Here we vary the interaction strength via the model parameter $g$ or
$m_{\sigma 8}$ in the interaction kernel~\eqref{eq:Vij}, and by
introducing the contribution from the $\pi N$ channel.  Since we are
interested in how the signal of the $\eprime n$ bound state depends on
the model parameters, we modify the interaction strength only for the
second scattering, i.e., $T_{2}$ in Fig.~\ref{fig:gamma-d}, while we
fix the first step of the reaction ($T_{1}$ in Fig.~\ref{fig:gamma-d})
unchanged.  We note that when we change the value of the parameter $g$
or $m_{\sigma 8}$, other parameters remain fixed as their original
values.

\begin{figure}[!t]
  \centering
  \Psfig{8.6cm}{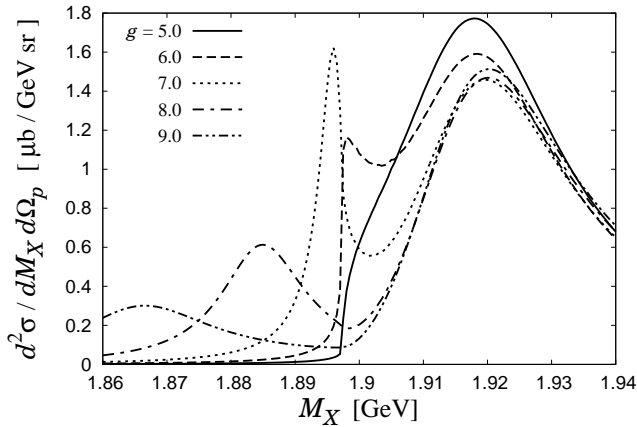} 
  \caption{Invariant mass spectrum for the $\gamma d \to X p$ reaction
    with $X = \eta n$ and $\eprime n$ as the sum of the two
    contributions.  The interaction strength is controlled by the
    parameter $g$ in Eq.~\eqref{eq:Vij}.  The initial photon energy is
    fixed as $\Eglab = 2.1 \gev$ and the proton scattering angle in
    the global center-of-mass is $\theta _{p}^{\rm cm}=0$ degree.}
  \label{fig:para_g}
\end{figure}

\begin{figure}[!b]
  \centering
  \Psfig{8.6cm}{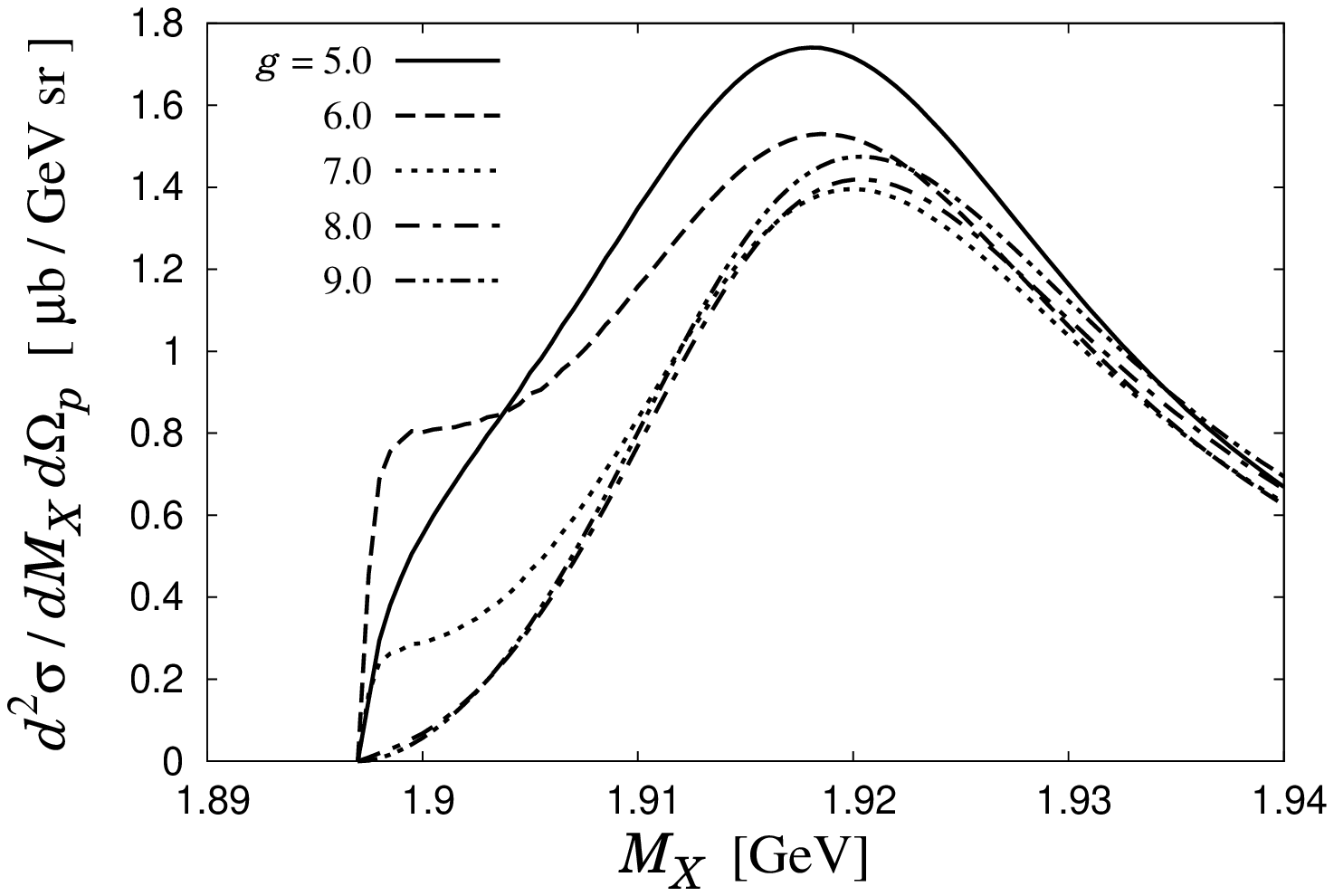} 
  \caption{Invariant mass spectrum for the $\gamma d \to \eprime n p$
    reaction.  The interaction strength is controlled by the parameter
    $g$ in Eq.~\eqref{eq:Vij}.  The initial photon energy is fixed as
    $\Eglab = 2.1 \gev$ and the proton scattering angle in the global
    center-of-mass is $\theta _{p}^{\rm cm}=0$ degree.}
  \label{fig:para_g_etaP}
\end{figure}

First we vary the interaction strength via $g$, which is the coupling
constant for the $\sigma N N$ vertex.  Since the coupling constant $g$
is commonly introduced to the $\eprime n \leftrightarrow \eprime n$
and $\eprime n \leftrightarrow \eta n$ interaction, as the value of
$g$ becomes large both the binding energy $B_{\rm E} \equiv
M_{\eprime} + M_{n} - \text{Re} \, w_{\rm pole}$ and width $\Gamma
\equiv - 2 \, \text{Im} \, w_{\rm pole}$ of the $\eprime n$ bound
state increase.  We show in the upper panel of Table~\ref{tab:2} the
properties of the $\eprime n$ bound state with several values of $g$.
We have checked that in the present condition the coupling constant
$g \ge 6.9$ can form an $\eprime n$ bound state below the $\eprime n$
threshold.

The behavior of the signal of the $\eprime n$ bound state is shown in
Fig.~\ref{fig:para_g}, where we plot the sum of the differential cross
sections of $\gamma d \to \eta n p$ and $\eprime n p$ with the
parameter $g = 5.0$ to $9.0$ in intervals of $1.0$.  From the figure,
we can clearly observe the signal of the $\eprime n$ bound state for
$g = 7.0$ and $8.0$.  However, for $g = 9.0$, the signal of the bound
state becomes weak due to its large decay width, $\Gamma = 26.0 \mev$.
In addition, for $g = 6.0$, we find only a cusp structure at the
$\eprime n$ threshold, as the interaction with $g = 6.0$ cannot bind
the $\eprime n$ system below the $\eprime n$ threshold.  Such a cusp
structure disappears when we take $g = 5.0$.  This result indicates
that, if the $\eprime N$ interaction is attractive enough, we have a
chance to observe some peculiar structure around the $\eprime n$
threshold, i.e., the bound state signal ($g = 7.0$, $8.0$ and $9.0$)
or a cusp of the differential cross section at the $\eprime n$
threshold ($g = 6.0$).  We also note that we may observe interesting
behavior in the $\eprime n$ invariant mass spectrum just above its
threshold, which reflects the physics below the $\eprime n$ threshold,
as seen in Fig.~\ref{fig:para_g_etaP}, where we plot only the $\eprime
n$ invariant mass spectrum.  In the present model, one finds that the
$\eprime n$ invariant mass spectrum is convex downward just above the
$\eprime n$ threshold for $g > 6$, in which there is a bound state
below the threshold, while it turns to be convex upward for $g \le 6$,
where there is no bound state.

\begin{figure}[!t]
  \centering
  \Psfig{8.6cm}{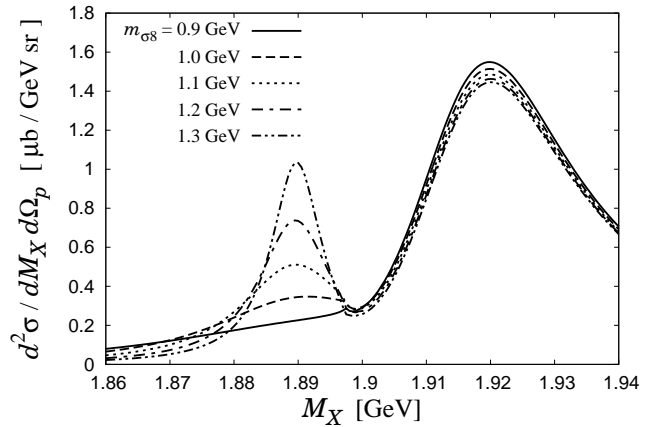} 
  \caption{Invariant mass spectrum for the $\gamma d \to X p$ reaction
    with $X = \eta n$ and $\eprime n$ as the sum of the two
    contributions.  The interaction strength is controlled by the
    parameter $m_{\sigma 8}$ in Eq.~\eqref{eq:Vij}.  The initial
    photon energy is fixed as $\Eglab = 2.1 \gev$ and the proton
    scattering angle in the global center-of-mass is $\theta _{p}^{\rm
      cm}=0$ degree.}
  \label{fig:para_Ms}
\end{figure}

Next, we shift the value of the parameter $m_{\sigma 8}$, which is the
mass of the octet $\sigma$ meson exchanged between $\eta ^{( \prime
  )}$ and $n$.  Since $m_{\sigma 8}$ determines the strength of the
transition $\eprime N \leftrightarrow \eta N$, this mainly controls
the decay width of the $\eprime n$ bound state; the smaller value of
$m_{\sigma 8}$ brings the larger decay width of the $\eprime n$ bound
state with a similar binding energy.  The properties of the $\eprime
n$ bound state are listed in the middle panel of Table~\ref{tab:2}.

By changing the value of $m_{\sigma 8}$, we can study how the bound
state signal melts with large decay width in the differential cross
section.  In Fig.~\ref{fig:para_Ms} we show our result of the sum of
the differential cross sections of $\gamma d \to \eta n p$ and
$\eprime n p$ with the parameter $m_{\sigma 8} = 0.9 \gev$ to $1.3
\gev$ in intervals of $0.1 \gev$.  We can see from
Fig.~\ref{fig:para_Ms} that for $m_{\sigma 8} \ge 1.1 \gev$ the signal
of the $\eprime n$ bound state is clear and nonnegligible compared to
the quasifree contribution above the $\eprime n$ threshold.  In
contrast, for $m_{\sigma 8} \le 1.0 \gev$, we have only negligible
contribution of the bound state signal.  This result indicates that,
even if there would exist an $\eprime n$ bound state, we could not see
its signal in the $\gamma d \to p n \eta$ reaction if its decay width
is $\Gamma \gtrsim 25 \mev$.

\begin{figure}[!t]
  \centering
  \Psfig{8.6cm}{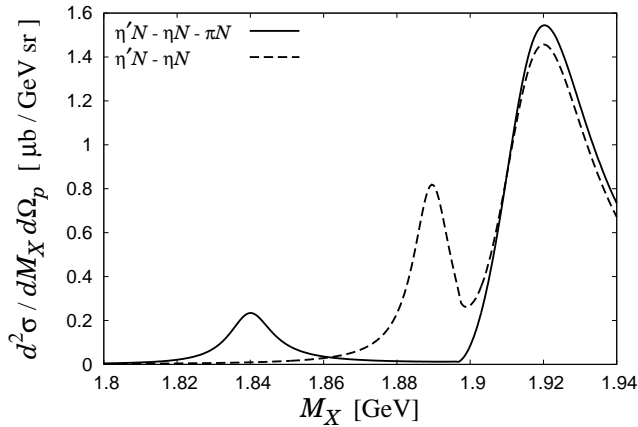} 
  \caption{Invariant mass spectrum for the $\gamma d \to X p$ reaction
    with $X = \eta n$ and $\eprime n$ as the sum of the two
    contributions.  Solid and dashed lines represent the invariant
    mass spectra with and without inclusion of the $\pi N$ channel,
    respectively.  The initial photon energy is fixed as $\Eglab = 2.1
    \gev$ and the proton scattering angle in the global center-of-mass
    is $\theta _{p}^{\rm cm}=0$ degree.}
  \label{fig:piN}
\end{figure}

Finally, we introduce the contribution from the $\pi N$ channel to the
$\eprime N$ interaction in the linear sigma model.  The contribution
from the $\pi N$ channel is included in order to respect the
experimental data given in Ref.~\cite{Rader:1973mx}.  Within this
treatment, the effect of the coupling with $\pi N$ channel would not
be so significant.  Besides, for a more realistic treatment of the
model, we also take into account the effect of the flavor SU(3)
symmetry breaking.  This SU(3) symmetry breaking makes the $\sigma
_{0}$ mass lighter.  As a result, the interaction in the $\eta
^{\prime} N$ elastic channel, which contains $m_{\sigma 0}$ in the
denominator, becomes more attractive and hence the binding energy of
the $\eta ^{\prime} N$ system increases.  In the present model, the
binding energy of the $\eprime N$ bound state grows to $57.0 \mev$,
which can be interpreted as a model parameter $m_{\sigma 0}$
dependence, but its decay width is still narrow, $14.5 \mev$.  The
details are given in Ref.~\cite{Sakai:2016}.  We note that, in the
calculation of the reaction cross sections, we do not take into
account the double scattering amplitude with the $\pi$ exchange, since
the exchanged $\pi$ should go far from its mass shell, which gives
only a negligible contribution.

We show in Fig.~\ref{fig:piN} the result of the sum of the
differential cross sections of $\gamma d \to \eta n p$ and $\eprime n
p$.  From the figure we can observe a clear signal of the $\eprime n$
bound state at $M_{X} = 1.84 \gev$ though the peak of the bound-state
signal is reduced compared with that without $\pi N$ channel.

\subsection{Behavior of the signal of the $\bm{\eprime n}$ bound state in
    several experimental conditions}

\label{sec:3-2}

Let us now discuss how the signal of the $\eprime n$ bound state can
be seen in several experimental conditions.  The model parameters are
the same as those given in Sec.~\ref{sec:2}.

\subsubsection{Photon energy dependence}

\begin{figure}[!t]
  \centering
  \Psfig{8.6cm}{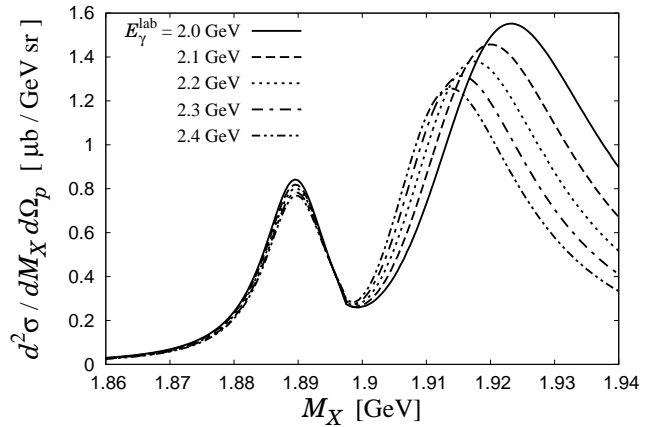} 
  \caption{Invariant mass spectrum for the $\gamma d \to X p$ reaction
    with $X = \eta n$ and $\eprime n$ as the sum of the two
    contributions.  The initial photon energy is taken from $\Eglab =
    2.0 \gev$ to $2.4 \gev$ in intervals of $0.1 \gev$.  The proton
    scattering angle in the global center-of-mass is $\theta _{p}^{\rm
      cm}=0$ degree.}
  \label{fig:Eglab}
\end{figure}

First we examine the initial photon energy dependence of the
differential cross section.  We take the initial photon energy from
$\Eglab = 2.0 \gev$ to $2.4 \gev$ in intervals of $0.1 \gev$ and the
proton scattering angle $\theta _{p}^{\rm cm} = 0$ degree.  The result
of the cross section around the $\eprime n$ threshold is plotted in
Fig.~\ref{fig:Eglab}.

From Fig.~\ref{fig:Eglab}, we can find that the peak height of the
signal of the $\eprime n$ bound state at $1.89 \gev$ is almost
unchanged as the initial photon energy increases.  This is due to the
two facts on the $\eprime$ photoproduction.  First, the $\gamma p \to
\eprime p$ reaction cross section, and hence its amplitude, decreases
as the photon energy increases, as seen in Fig.~\ref{fig:data}.
Second, with forward proton emission, $\eprime$ produced on a bound
proton becomes slower in the laboratory frame as the photon energy
increases, which makes the intermediate $\eprime$ close to on its mass
shell in the $\eprime n$ signal region, and hence the $\eprime$
exchange contribution becomes stronger.  These two contributions
compensate each other, and as a result the signal of the bound state
is almost unchanged regardless of the initial photon energy.  On the
other hand, while the peak height of the $\eprime$ quasifree
contribution seen above the $\eprime n$ threshold is similar, its peak
position shifts downward as the photon energy increases.  This is
caused by that $\eprime$ produced on a bound proton becomes slower in
the laboratory frame as the photon energy increases, which makes the
$\eprime n$ invariant mass lower.

\subsubsection{Scattering angle dependence}

\begin{figure}[!t]
  \centering
  \Psfig{8.6cm}{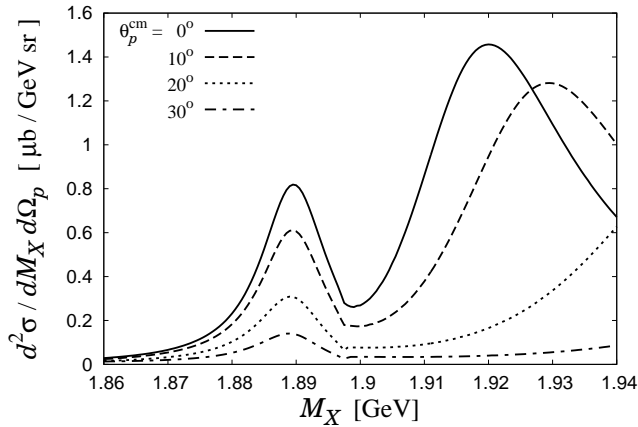} 
  \caption{Invariant mass spectrum for the $\gamma d \to X p$ reaction
    with $X = \eta n$ and $\eprime n$ as the sum of the two
    contributions.  The scattering angle of the final-state proton is
    taken from $\theta _{p}^{\rm cm} = 0$ to $30$ degrees in the global
    center-of-mass frame.  The initial photon energy is fixed as
    $\Eglab = 2.1 \gev$.}
  \label{fig:angle}
\end{figure}

Next, we change the value of the scattering angle of the final-state
proton.  Here we take the scattering angle in the global
center-of-mass frame, $\theta _{p}^{\rm cm}$, from $0$ to $30$ degrees
in intervals of $10$ degrees.  The result of the differential cross
section in these values of the scattering angle is shown in
Fig.~\ref{fig:angle}.  From the figure, for larger scattering angle
$\theta _{p}^{\rm cm}$, we observe smaller bound state signal.  This
is because, with finite $\theta _{p}^{\rm cm}$, exchanged $\eprime$
goes largely off-shell due to a large transverse momentum and hence
the $\eprime$ exchange contribution becomes weak.  Therefore, this
result indicates that the forward proton emission is suitable for the
production of the $\eprime n$ bound state, as we have expected.
However, we also see that the quasifree $\eprime$ peak shifts upward
due to the same kinematics.  This fact may help us to observe the
signal of the $\eprime n$ bound state in actual experiments, as in
experiments we measure the production cross sections with finite
scattering angles.

\subsubsection{Integrating the angle for forward proton emission}

\begin{figure}[!t]
  \centering
  \Psfig{8.6cm}{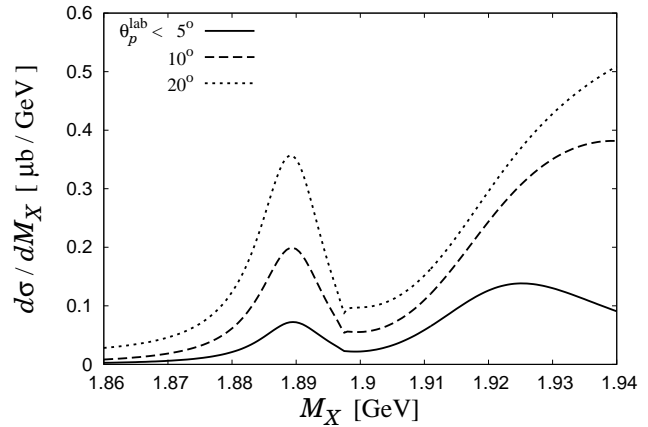} 
  \caption{Invariant mass spectrum for the $\gamma d \to X p$ reaction
    with $X = \eta n$ and $\eprime n$ as the sum of the two
    contributions.  The scattering angle of the final-state proton in
    the laboratory frame, $\theta _{p}^{\rm lab}$, is integrated in
    the ranges [$0^{\circ}$, $5^{\circ}$], [$0^{\circ}$,
      $10^{\circ}$], and [$0^{\circ}$, $20^{\circ}$].  The initial
    photon energy is fixed as $\Eglab = 2.1 \gev$.}
  \label{fig:lab}
\end{figure}

Finally, in order to see the cross section corresponding to the
realistic experimental observations, we show the cross section
integrated with respect to the scattering angle for forward proton
emission in the laboratory frame in Fig.~\ref{fig:lab}.  The result
indicates that, in any cases of the upper limit of the scattering
angle, we can clearly distinguish the signal of the $\eprime n$ bound
state, if existed, from the $\eprime$ quasifree contribution.  This
result indicates that we will observe the signal of the $\eprime n$
bound state in experiments of the $\gamma d \to \eta ^{( \prime )} n
p$ reaction with forward proton emission, especially if the bound
state exists at more than several MeV below the $\eprime n$ threshold
with a small decay width.

\section{Conclusion}
\label{sec:4}

In this study, we have investigated possibilities of observing a
signal of an $\eprime n$ bound state in the photoproductions of the
$\eta$ and $\eprime$ mesons on a deuteron target with forward proton
emission.  For this purpose, we have described the production process
by two portions.  One is the photoproduction of the $\eta ^{( \prime
  )}$ meson on a proton, and the other is the $\eta ^{( \prime )} n
\to \eta ^{( \prime )} n$ scattering.  In this study, the $\eta ^{(
  \prime )} N \to \eta ^{( \prime )} N$ interaction is obtained in the
linear sigma model, and this interaction is employed as a kernel of
the scattering equation so as to calculate the $s$-wave $\eta ^{(
  \prime )} N$ scattering amplitude, in which an $\eprime N$ bound
state can be dynamically generated.  On the other hand, the $\gamma p
\to \eta p$ and $\eprime p$ scattering amplitudes are fixed in an
effective model so as to reproduce the experimental cross sections
with forward proton emission.

By using these two portions, we have calculated cross sections of the
$\gamma d \to \eta n p$ and $\eprime n p$ reactions with forward
proton emission in single and $\eta ^{(\prime )}$-exchange double
scattering processes.  As a result, we have found that the signal of
the $\eprime n$ bound state can be seen below the $\eprime n$
threshold in the $\eta n$ invariant mass spectrum of the $\gamma d \to
\eta n p$ reaction and its strength is comparable with the
contribution from the quasifree $\eprime$ production above the
$\eprime n$ threshold in the $\eprime n$ invariant mass spectrum.  We
have found that the double scattering process of the $\eprime$
exchange dominates the production of the $\eprime n$ bound state.  We
have also seen a nonnegligible destructive interference between the
$\eprime$ quasifree contribution in the single scattering and the tail
of an $\eprime n$ bound state peak coming from the double scattering
of the $\eprime$ exchange, due to the $\eprime$ absorption into the
bound neutron.  Changing the strength of the $\eprime n$ interaction,
we have obtained a clear signal of the $\eprime n$ bound state if its
decay width is about $10 \mev$.  In considering realistic experimental
conditions such as several initial photon energy and scattering angle,
we have concluded that we will observe the signal of the $\eprime n$
bound state in experiments of the $\gamma d \to \eta ^{( \prime )} n
p$ reaction with forward proton emission, especially in the case that
the bound state exists at more than several MeV below the $\eprime n$
threshold with a small decay width.

\begin{acknowledgments}
  The authors thank N.~Muramatsu, M.~Sumihama, and T.~Ishikawa for
  useful discussions.
  T.~S.\ acknowledges the support by the Grants-in-Aid for young
  scientists from JSPS (No.~15K17649) and for JSPS fellows
  (No.~15J06538).  S.~S.\ was a JSPS fellow and appreciates the
  support of a JSPS Grant-in-Aid (No.~25-1879). The work of D.~J.\ was
  partly supported by Grants-in-Aid for Scientific Research from JSPS
  (25400254).
\end{acknowledgments}

\end{document}